\documentclass{article}
\usepackage[T1]{fontenc}
\usepackage{graphicx}
\usepackage{multirow}
\usepackage{hyperref}
\usepackage{makecell}
\usepackage{pdfpages}
\usepackage{tabularx}
\usepackage{paralist}
\usepackage{authblk}
\newcolumntype{L}[1]{>{\raggedright\arraybackslash}p{#1}}
\newcolumntype{C}[1]{>{\centering\arraybackslash}p{#1}}
\newcolumntype{R}[1]{>{\raggedleft\arraybackslash}p{#1}}

\begin{document}

% macros
\newcommand{\Dataset}[1][]{D_{#1}}

\title{Simple Domain Adaptation for Sparse Retrievers}

\author[1,2]{Mathias Vast} 
\author[2]{Yuxuan Zong}
\author[1]{Basile Van Cooten}
\author[2]{ Benjamin Piwowarski} 
\author[2]{Laure Soulier}

\affil[1]{Sinequa}
\affil[2]{Sorbonne Université, CNRS, ISIR, F-75005 Paris, France \\ \{firstname.lastname\}@isir.upmc.fr}

\maketitle              % typeset the header of the contribution
\begin{abstract}
In Information Retrieval, and more generally in Natural Language Processing, adapting models to specific domains is conducted through fine-tuning. Despite the successes achieved by this method and its versatility, the need for human-curated and labeled data makes it impractical to transfer to new tasks, domains, and/or languages when training data doesn't exist. Using the model without training (zero-shot) is another option that however suffers an effectiveness cost, especially in the case of first-stage retrievers.
Numerous research directions have emerged to tackle these issues, most of them in the context of adapting to a task or a language. However, the literature is scarcer for domain (or topic) adaptation.
In this paper, we address this issue of cross-topic discrepancy for a sparse first-stage retriever by transposing a method initially designed for language adaptation. By leveraging pre-training on the target data to learn domain-specific knowledge, this technique alleviates the need for annotated data and expands the scope of domain adaptation. Despite their relatively good generalization ability, we show that even sparse retrievers can benefit from our simple domain adaptation method.
\end{abstract}
\section{Introduction}

Nowadays, most of the models that achieve state-of-the-art results in Natural Language Processing (NLP) rely on the "\textit{pre-train then fine-tune}" pipeline~\cite{bert,monobert}. In this setup, a model is first trained with self-supervised objectives, such as  Masked Language Modeling, on a large unlabeled corpus, before being fine-tuned on a labeled dataset for a specific task. In Information Retrieval (IR), this method has delivered huge improvements over the previous "pre-BERT" models and has given birth to a large variety of frameworks including, but not limited to, dense retrievers \cite{karpukhin-etal-2020-dense}, learned sparse retrievers \cite{Formal_Lassance_Piwowarski_Clinchant_2021} and cross-encoders \cite{monobert}. For a given task, when enough good-quality labeled data is available, fine-tuning an already pre-trained model is the best option. Thanks to datasets such as MS MARCO \cite{MSMARCO}, this condition might be met for the majority of the downstream tasks in the generic domain for the English language, but in other languages and/or specific domains, these resources may not exist. 

This poses a problem since a fine-tuned model evaluated outside its training domain might perform poorly~\cite{thakur_beir_2021}. When the volume of training data is insufficient for fine-tuning, it can even degrade the model's performance~\cite{little_worse_than_none}.  Besides, the increasing size of these models coupled with the difficulty of collecting a satisfying amount of high-quality labeled data has led to a growing interest in zero-shot or few-shot methods. 

It is possible to tackle this issue by leveraging unlabeled data and pursuing pre-training over task or domain-related data~\cite{domain-adaptive_pretraining,task-adaptive_pretraining} before fine-tuning. Providing a better initialization point alleviates the dependency on the amount of in-domain labeled data available for fine-tuning. Differently, Artetxe et al.~\cite{Artetxe_Ruder_Yogatama_2020} leverage pre-training as a means to adapt a model \emph{subpart} for a specific language, different from the source language used for pre-training and fine-tuning the base model. 
In this paper, we consider a setup of cross-domain adaptation where the model to be transferred can be trained on a source domain $\Dataset[source]$, where an annotated corpus containing enough documents for both pre-training and fine-tuning is available, whereas the target domain $\Dataset[target]$ contains enough data for pre-training but no annotation.
We extend the work of~\cite{Artetxe_Ruder_Yogatama_2020} in two ways to deal with ad-hoc IR. 
First, inspired by~\cite{clark2019attention}, we study which subparts of the model should be pre-trained or fine-tuned -- we do not only consider the embeddings to be domain-dependent. 
% 
% Second, as we are not considering cross-lingual adaptation anymore but the base model still needs prior knowledge about the language, we propose a new pre-training procedure on both the source and target domains. Our procedure disentangles the pre-training on the source dataset from the learning of the base knowledge required by the model and allows to use Artetxe et al.'s pipeline outside its original scope.
Second, we propose a new pre-training procedure on both the source and target domains, and not only on the target one. This allows a model to learn task-specific parameters (on the source domain) when (and only when) its domain-specific ones are well set.
As domain shift impacts more strongly first-stage retrievers, we study a first-stage sparse retriever, namely SPLADE~\cite{Formal_Lassance_Piwowarski_Clinchant_2021}. Our approach allows one to share the fine-tuning resources across multiple domains with less drop in effectiveness due to domain discrepancies. 

%\vspace{-0,2cm}
\section{Related works}
\label{related_works}

Transfer learning and adaptation to a specific context are long-standing research topics. With the recent arrival of large language models, research shifted to transferring these models at the lowest computational and annotation cost possible~\cite{budget_aware}. To get rid of human annotation, works inspired by Doc2Query~\cite{doc2query} propose to generate queries for which a document in the target domain would be relevant~\cite{li2022domain,gpl}. Inversely, generative models can also be used to produce pseudo-relevant documents~\cite{gao2022precise} before fine-tuning a model on it. \cite{hashemi-etal-2023} even proposed a method that generates a complete collection of both documents and queries from a simple description of the target domain. For each of these methods, one can apply the traditional "\textit{pretrain then fine-tune}" framework. Besides the fact that this type of approach can be computationally costly, studying, as in our work, pre-training techniques, is complementary. Indeed, as highlighted in \cite{experimental_study_on_pretraining_2023}, better pre-training, either by using different self-supervised objectives \cite{izacard_unsupervised_2022} or pursuing it longer \cite{domain-adaptive_pretraining,task-adaptive_pretraining}, often provides better results compared to fine-tuning for first-stage methods.

% Because of the cost of pre-training a language model, 
Other works have tried to ease the transfer of (large) pre-trained models. We can separate them into two categories, even though they share the same underlying principle of distinguishing different sets of parameters during training. The first family, referred to as Parameter-Efficient Fine-Tuning (or PEFT) \cite{houlsby_parameter-efficient_2019,hu_lora_2021,li_prefix-tuning_2021,liu_p-tuning_2022}, achieves this objective by modifying the original model with an \emph{adapter} whose parameters are fine-tuned. Its success relies on the quality of the underlying pre-trained model, whose generality is worth being conserved while tuning only a small portion of parameters on the target task. 
% Previous studies had highlighted the possibility of achieving performances on par with full fine-tuning while only tuning some layers~\cite{layers_freezing}. 
In IR, these methods have started to be explored recently and, so far, have proven to be successful \cite{peft_for_ranking,Pal_Lassance_Dejean_Clinchant_2023,tam_parameter-efficient_2022}. Orthogonal to these works, some papers do not use freezing to preserve generality or to spare computations, but to adapt specific parts of the model for a given language or task. In particular, \cite{zhan_disentangled_2022} applies PEFT methods to Dense Retrievers to disentangle domain and relevance learning. For each domain, a Transformer backbone is trained to learn its linguistic features~(MLM). Separately, a distinct module is trained once and then shared, to predict relevance based on these extracted features. Artetxe et al. \cite{Artetxe_Ruder_Yogatama_2020} adapt a similar reasoning but apply the disentangling within the Transformer backbone's layers, thus introducing a zero inference overhead. In this work, we extend their approach to domain adaptation in the context of IR. Note that using PEFT following~\cite{zhan_disentangled_2022} is an alternative that we will explore in the future. 

\section{Cross-Domain Adaptation of a Neural Network model}
\label{proposed_method}

\begin{figure}[t] % <---
    \includegraphics[width=\textwidth]{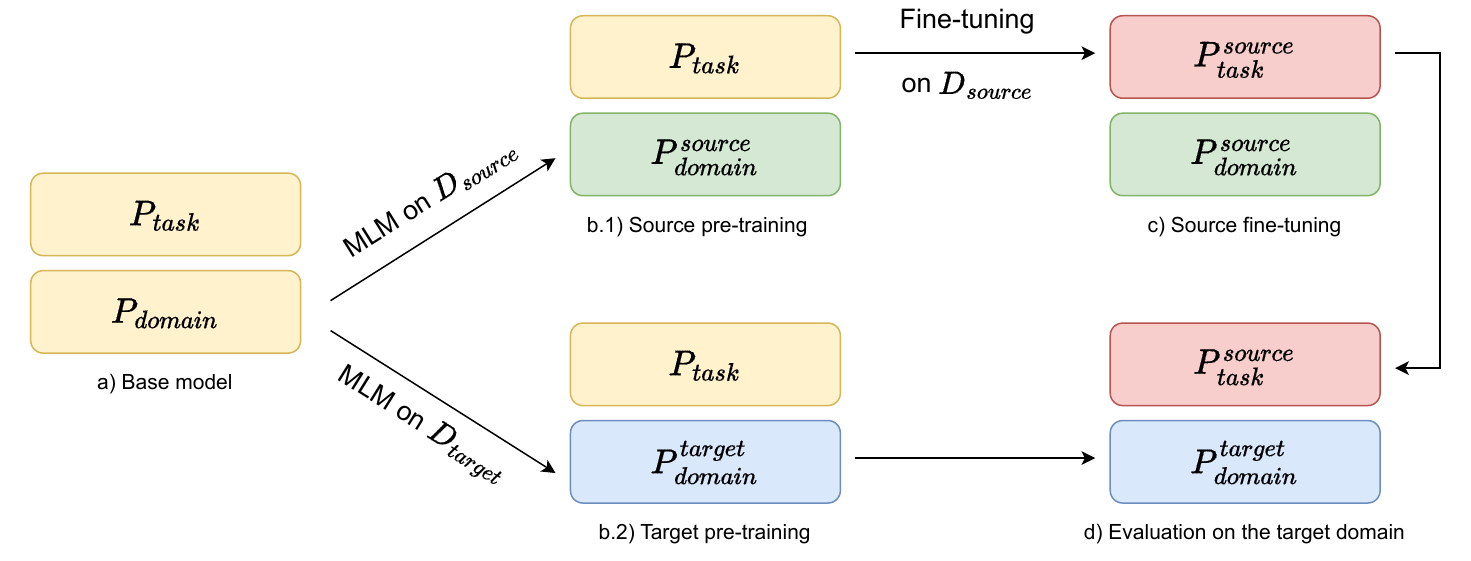}
   \caption{Illustration of the cross-domain adaptation process} % Need to check a little about this
   \label{fig:pipeline}
   \vspace{-0.5cm}
\end{figure}

We describe here how we adapted the pipeline introduced by Artetxe et al. \cite{Artetxe_Ruder_Yogatama_2020} to cross-domain adaptation. This approach can be extended to any model in any setting by establishing a distinction between the model's parameters dedicated to learning the domain/language model, denoted $P_{domain}$, and the ones dedicated to learning the task denoted $P_{task}$, which are supposed to be distinct, i.e. $P_{domain} \cap P_{task} = \emptyset$. Given the IR setting, we define $P_{domain}$ to be the embeddings together with the first $k$ layers of the transformer architecture -- whereas in the language adaptation setting, \cite{Artetxe_Ruder_Yogatama_2020} only considered $P_{domain}$ to be the embeddings. 

Figure \ref{fig:pipeline} describes the training process made of two distinct pre-trainings; source (b.1) and target (b.2); and one fine-tuning on the source (c), before evaluating (d) on the target domain:
\begin{description}
    \item[Pre-training] for $\Dataset[source]$ (resp. $\Dataset[target]$): As in~\cite{Artetxe_Ruder_Yogatama_2020}, we continue to pre-train BERT over $\Dataset[source]$ (resp. $\Dataset[target]$) while keeping the parameters' subset $P_{task}$ frozen. This step adapts the parameters $P_{domain}$ to the specificity of the domain $\Dataset[source]$ (resp. $\Dataset[target]$). We denote $P_{domain}^{source}$ (resp. $P_{domain}^{target}$) the parameters after the MLM pre-training. Note that contrary to ~\cite{Artetxe_Ruder_Yogatama_2020}, we keep the base model's vocabulary unchanged throughout the pipeline.
    \item[Fine-tuning] on $\Dataset[source]$: We leverage the previously pre-trained subset $P_{domain}^{source}$ but keep it frozen while only fine-tuning the $P_{task}$ parameters. The objective of this step is to specialize the subset of parameters $P_{task}$ to the IR task. 
    We denote as $P_{task}^{source}$ the fine-tuned parameters.
\end{description}

Contrary to~\cite{Artetxe_Ruder_Yogatama_2020}, we add an additional pre-training step over the source corpus (b.1 in Figure \ref{fig:pipeline}). We suppose that thanks to this step, the model can better learn the downstream task as it is not fine-tuned on the same dataset (MS MARCO) that it was originally pre-trained on (Wikipedia). 

Finally, for inference on the $\Dataset[target]$'s dataset, we combine the two subsets $P_{domain}^{target}$ and $P_{task}^{source}$. The underlying intuition is that the subset $P_{task}^{source}$ is task-specific while $P_{domain}^{target}$ is domain-specific. The model should be able to generalize better when used forIR in the target domain.

\section{Experiments}
\label{experiments}

We use $BERT_{base}$ pre-trained on the English version of Wikipedia\footnote[2]{The model is made available by Google on the HuggingFace' Hub: \href{https://huggingface.co/bert-base-uncased}{bert-base-uncased}} as our base model and share its base vocabulary through every domain. We focus on a sparse first-stage ranker, SPLADE~\cite{Formal_Lassance_Piwowarski_Clinchant_2021}, as first-stage rankers particularly suffer from domain shifts and SPLADE is known to be at the state-of-the-art among them.

\subsection{Datasets, baselines, and ablations}

% Describe the dataset we evaluate, including their domain, type, etc
Two datasets in IR might differ in 1) domain/vocabulary, 2) document types, and/or 3) topic types \cite{Pan_survey_transfer_2010}. For a neural model, any of these distinctions can impact the generalization power.
The source dataset we consider across all our experiments is MS MARCO \cite{MSMARCO} as it is the defacto dataset for fine-tuning pre-trained models in IR and its domain can be regarded as "generic Web search". We assemble a collection of target datasets from a subset of the BEIR benchmark (*) \cite{thakur_beir_2021}, the LoTTE's collection (**) \cite{ColBERTv2_2022}, as well as other sources. The decision about using or not some datasets from the BEIR benchmark is based on multiple criteria, including the existence of distinct train and test sets and the task affiliated with the dataset as we are only interested in pure IR in this case. We processed them with the \textit{ir-datasets} library \cite{macavaney:sigir2021-irds}. Table \ref{tab:datasets} gives details about the characteristics of each dataset and shows how we covered various domains, ranging from bio-medical to literature.

\begin{table}[tb]
\caption{Datasets considered in the study. Statistics on the average topic and document length for each dataset are computed using BERT tokens.}\label{tab:datasets}
\begin{center}
\begin{tabular}{|c|c|c|c|c|}
\hline
\multirow{2}{*}{Dataset} & \multirow{2}{*}{Domain} & \multicolumn{3}{c|}{Dataset Statistics} \\
\cline{3-5}
& & Avg top./doc. len & \# Top & \# Doc \\
\hline
\hline
MSM-Passage & Generic & 7.5 / 74.9 & 808K & 8.8M \\
\hline
TREC-COVID (*) & Bio-Medical & 16.0 / 243.5 & 50 & 171.3K \\
NFCorpus (*) & Bio-Medical & 5.0 / 338.1 & 325 & 5.4K \\
BioASQ (*) & Bio-Medical & 13.1 / 320.3 & 500 & 14.9M \\
FiQA-2018 (*) & Finance & 13.6 / 175.1 & 648 & 57.6K \\
TREC-NEWS (*) & News & 18.2 / 652.5 & 50 & 595K \\
Robust04 (*) & News & 18.7 / 638.5 & 250 & 528.1K \\
ANTIQUE \cite{antique} & Web & 11.9 / 52.3 & 200 & 403.7K \\
LoTTe-Wri. (**) & Literature & 8.7 / 165.2 & 1071 & 200K \\
LoTTe-Tec. (**) & Technology & 9.7 / 216.6 & 596 & 638.5K \\
LoTTe-Rec. (**) & Entertainment & 9.3 / 181.6 & 924 & 167K \\
\hline
\end{tabular}
\end{center}
%\vspace{-0.2cm}
\end{table}

% Need to recheck the t-test values
\begin{table}[tb]
  \caption{Performance in nDCG@10 of our approach versus BM25 and Zero-shot (SPLADE). Bold values are strictly superior to the Zero-shot baseline. $\dagger$ and $\ddagger$: Improves upon BM25 baseline and Zero-Shot baseline respectively with statistical significance ($p\leq0.05$) under the two-tailed Student’s t-test.}
   \label{tab:results_splade}
   \resizebox{\textwidth}{!}{
    \begin{tabular}{|c|cc|C{1.1cm}C{1.1cm}C{1.1cm}C{1.1cm}|cc|}
    \hline
    \textbf{Method} ($\rightarrow$) & \multicolumn{2}{c|}{Baselines} & \multicolumn{4}{c|}{Proposed pipeline ($k=\ldots$)} & \multicolumn{2}{c|}{Ablations} \\
    \hline
    \textbf{Dataset} ($\downarrow$) & BM25 & 0-shot & 0 layer & 1 layer & 2 layers & 4 layers & w/o pre-training & w/o source\\
    \hline
    TREC-COVID & 58.1 & 71.3 & $68.8^\dagger$ & $70.5^\dagger$ & $\textbf{72.1}^\dagger$ & $70.9^\dagger$ & $67.9^\dagger$ & $67.7^\dagger$ \\\hline
    NFCorpus & 24.5 & 24.9 & $\textbf{25.6}$ & $\textbf{25.4}$ & $\textbf{25.1}$ & 24.1 & $\textbf{25.1}$ & $\textbf{25.6}$\\\hline
    BioASQ & \textbf{52.3} & 43.1 & $\textbf{45.4}^\ddagger$ & \textbf{44.2} & $\textbf{45.2}^\ddagger$ & $\textbf{46.2}^\ddagger$ & $\textbf{44.8}^\ddagger$ & $\textbf{45.3}^\ddagger$\\\hline
    FiQA-2018 & 23.6 & 27.3 & $27.2^\dagger$ & $\textbf{27.5}^\dagger$ & $\textbf{29.5}^{\dagger\ddagger}$ & $\textbf{29.0}^{\dagger\ddagger}$ & $25.2^\dagger$ & $\textbf{28.3}^\dagger$\\\hline
    TREC-NEWS & 31.4 & 32.1 & \textbf{33.3} & \textbf{32.8} & \textbf{33.0} & $\textbf{36.5}^{\ddagger}$ & 31.9 & \textbf{33.1}\\\hline
    Robust04 & 40.8 & 39.6 & $\textbf{42.9}^\ddagger$ & $\textbf{42.9}^\ddagger$ & $\textbf{43.5}^\ddagger$ & $\textbf{42.6}^\ddagger$ & \textbf{41.2} & $\textbf{42.1}^\ddagger$\\\hline
    ANTIQUE & \textbf{45.4} & 43.3 & 42.9 & \textbf{44.0} & \textbf{43.7} & \textbf{44.1} & $\textbf{45.0}^\ddagger$ & $\textbf{45.0}^\ddagger$\\\hline
    LoTTE-Rec. & 39.3 & 46.2 & $\textbf{47.6}^{\dagger\ddagger}$ & $\textbf{47.7}^{\dagger\ddagger}$ & $\textbf{47.8}^{\dagger\ddagger}$ & $\textbf{46.9}^\dagger$ & $\textbf{46.8}^\dagger$ & $\textbf{46.7}^\dagger$\\\hline
    LoTTE-Wri. & 41.3 & 52.7 & $\textbf{53.0}^\dagger$ & $\textbf{53.8}^{\dagger\ddagger}$ & $\textbf{53.4}^\dagger$ & $\textbf{54.1}^{\dagger\ddagger}$ & $52.0^\dagger$ & $\textbf{52.9}^\dagger$\\\hline
    LoTTE-Tec. & 25.3 & 36.8 & $\textbf{38.2}^\dagger$ & $\textbf{37.6}^\dagger$ & $\textbf{37.4}^\dagger$ & $36.8^\dagger$ & $36.8^\dagger$ & $36.8^\dagger$\\\hline\hline
    Avg. & 38.2 & 41.7 & \textbf{42.5} & \textbf{42.6} & \textbf{43.1} & \textbf{43.1} & \textbf{41.9} & \textbf{42.4}\\\hline
    \end{tabular}
    }
\end{table}

% \subsection{Baselines and variants}

% Describe the baselines, including the difference from our proposed methods and the hyperparameters to be used
To verify the benefits of our adaptation method, we experimented with multiple variants, differing on the $P_{domain}$ / $P_{task}$ split by the number of layers $k \in \{0, 1, 2, 3, 4, 6, 8, 10\}$ after the embedding belonging to $P_{domain}$. We also included two baselines and two ablations that we describe in what follows:
\begin{itemize}
    \item \textbf{BM25} \cite{bm25}, is a very strong baseline in domain adaptation \cite{thakur_beir_2021} and is also a sparse first-stage retriever;
    \item \textbf{Zero-Shot Learning} evaluation of the SPLADE model fine-tuned on MS MARCO to quantify the benefits we can obtain with our adaptation method;
    \item Ablation of the source pre-training, referred to as \textbf{w/o source}, as this corresponds to the original approach described in Artetxe et al. (with $k=0$);
    \item Ablation of the source \emph{and} target pre-trainings, referred to as \textbf{w/o pre-training} (again with $k=0$). Note that this is a zero-shot model. As we are starting from a model pre-trained on Wikipedia, this ablation evaluates whether our proposition has an impact.
\end{itemize}
All the project code, based on the library \textit{experimaestro-ir} \cite{xpmir}, including the experimental details, is freely accessible\footnote{\href{https://git.isir.upmc.fr/mat\_vast/cross\_domain\_adaptation}{\url{https://git.isir.upmc.fr/mat\_vast/cross\_domain\_adaptation}}}.

\subsection{Results}

Table \ref{tab:results_splade} contains the results of these methods from which we can draw the following conclusions:
% \vspace{-0,2cm}
\begin{inparaenum}[(i)]
    \item Compared to the Zero-Shot Learning baseline, our adaptation approach allows us to gain on average between 0.7 and 1.4 points in nDCG@10, and more than 1 point for some datasets;
    \item Comparing "w/o source" to our approach, it seems that the additional pre-training step over the source domain helps the model to generalize better, with an average improvement between 0.1 and 0.8 points (for nDCG@10). This is sensible since the model learns to adapt the task-specific parameters from domain-specific parameters;
    \item Similarly, no pre-training at all can hurt the performance up to 1.2 points in nDCG@10 ("w/o pre-training"), showing the benefit of our pre-training procedure. However, in a zero-shot setting, we observe that fine-tuning only the Transformer's layers performs ("w/o pre-training") on par with fine-tuning the whole model ("0-shot"). Both setups are evaluated without using the target dataset, and can thus be considered as Zero-Shot Learning. It indicates that we could save some computations during fine-tuning without hurting performance.
    \item Pre-training over additional layers can provide additional gains of up to 0.5 points in nDCG@10 (columns "1 layer", "2 layers" and "4 layers" compared to the column "0 layer"). Results however seemed to plateau beyond $k=2$ as illustrated by the minor differences between the columns $k=4$, with the best result in average, and $k=2$. Table \ref{perf_and_sparsity} summarizes the evolution of the average performance along with the number of additional layers reserved for learning the target domain. 
    These results seem to comfort recent findings on the role of the first layers' attention head in a Transformer model~\cite{clark2019attention}. It also highlights that even though SPLADE-base models highly rely on the Masked Language Modeling task, fine-tuning nevertheless remains an important part of the process, and increasing the share of the parameters dedicated to the former can infringe the model's performance on the IR task. 
\end{inparaenum}

% \begin{figure}[t] %
%     \includegraphics[width=\textwidth]{figures/performance.png}
%    \caption{Evolution of the performance and of the sparsity of SPLADE when the number of layers dedicated to the learning of the target domain increases. Red curve describes the average performance evolution and blue curve the sparsity of the document encoder.} 
%    \label{fig:perf_and_sparsity}
%    \vspace{-0.5cm}
% \end{figure}
\begin{table}[tb]
\caption{Evolution of the performance in nDCG@10 and of the sparsity of SPLADE when the number of layers dedicated to the learning of the target domain increases. Best result is highlighted in bold.}
\label{perf_and_sparsity}
 \resizebox{\textwidth}{!}{
\begin{tabular}{|c|C{0.9cm}|C{0.9cm}|C{0.9cm}|C{0.9cm}|C{0.9cm}|C{0.9cm}|C{0.9cm}|C{0.9cm}|C{0.9cm}|}
\hline
\textbf{Method} ($\rightarrow$) & 0-shot & k=0  & k=1  & k=2  & k=3  & k=4  & k=6  & k=8  & k=10 \\ \hline
\begin{tabular}[c]{@{}l@{}}Average Performance \\ (nDCG@10)\end{tabular} & 41.7 & 42.5 & 42.6 & 43.1 & 42.6 & \textbf{43.1} & 43.1 & 43.0 & 42.6 \\ \hline
Sparsity (x$10^3$) & 9.3 & 7.3  & 9.2  & 11.5 & 8.4  & 10.3 & 10   & 9.8  & 9.3 \\ \hline
\end{tabular}
}

\end{table}

Table \ref{perf_and_sparsity} also gives additional insights into the sparsity achieved by the document encoder of our SPLADE variants. Query encoder sparsity isn't specified as it says consistent along the variants and the baseline. We note that the best models usually have a higher sparsity level, which tends to indicate that the associated model is able to build a more appropriate representation of each target domain. However, more work is needed to understand why some other variants happen to have lower sparsity values than the baseline while performing better.

\paragraph{Discussion and limitations.}
This work is a preliminary study of how pre-training can be used to ease domain adaptation in IR.
We led the same series of experiments with a second-stage ranker, namely monoBERT~\cite{monobert}. Interestingly, the results showed that our approach didn't provide any improvements compared to the Zero-Shot Learning approach in this case. Previous work \cite{experimental_study_on_pretraining_2023} already mentioned the fact that second-stage rankers benefit more from larger collections compared to pre-training on specific datasets. Our conclusion is that our approach is too naive to provide any improvement at all, given the generalization power of monoBERT, and suggests that a better understanding of which parameters are domain or task-specific, as well as their interplay, is necessary. Further work is also needed to explore the impact of pre-training time given corpus length on the final results. We leave both of these directions for future works as well as the understanding of why second-stage rankers cannot draw any benefits from specific pre-training.
Finally, we did not compare with generative methods (i.e., generating a query matching a document in the target domain) such as~\cite{li2022domain} -- these methods are costly, and we also need to investigate whether the improvements are complementary with our (more cost-efficient) approach.

\section{Conclusion}
\label{conclusion}

 We presented a simple approach to domain adaptation for IR, based on a language adaptation approach~\cite{Artetxe_Ruder_Yogatama_2020}. Results from our experiments highlight the potential of our method with learned sparse retrievers and could be particularly helpful when transferring fine-tuned models in contexts where high-quality annotated data isn't available or impossible to collect because it is either too expensive or complex. Another benefit of our approach is its re-usability. Once the expensive fine-tuning has been performed, the subset $P_{task}$ can be re-used with different subsets $P_{domain}$ pre-trained over different domains, which eliminates the need to perform a fine-tuning over all the model for every new domain. In the future, we would like to include dense retrievers as well as more costly methods in our study. In addition, we also want to extend it to more sophisticated pre-training approaches.

\section{Ackowledgements}
\label{acknwoledgements}
This work is supported by the ANR project GUIDANCE ANR-23-IAS1-0003. \newline
This work was granted access to the HPC resources of IDRIS under the allocation 2023-AD011014444 made by GENCI.
\bibliographystyle{splncs04}
\bibliography{biblio}

% \printbibliography

\end{document}